\documentstyle[prl,aps,epsf,epsfig,floats,twocolumn]{revtex}
\begin{document}
\draft
\twocolumn[\hsize\textwidth\columnwidth\hsize\csname
@twocolumnfalse\endcsname

\title{
\hfill{\small{FZJ-IKP(TH)-2000-05}}\\[0.6cm]
Complete one--loop analysis of the nucleon's spin polarizabilities}
\author{George C. Gellas, Thomas R. Hemmert and Ulf-G. Mei{\ss}ner}
\address{Forschungszentrum J{\"u}lich, Institut f{\"u}r Kernphysik (Theorie),
D-52425 J{\"u}lich, Germany}
\maketitle
\begin{abstract}
We present a complete one--loop analysis of the four nucleon spin
polarizabilities in the framework of heavy baryon chiral perturbation
theory. The first non--vanishing contributions to the
isovector and first corrections to the isoscalar spin polarizabilities
are calculated. No unknown parameters enter these predictions. We
compare our results to various dispersive analyses. We also
discuss the convergence of the chiral expansion and the role of the
delta isobar.
\end{abstract}

]

\vspace{1cm}

Low energy Compton scattering off the nucleon is an important
probe to unravel the nonperturbative structure of QCD since the electromagnetic
interactions in the initial and final state are well understood.
In the long wavelength
limit only the charge of the target can be detected and the experimental cross
sections, up to photon energies $\omega$ of about 50~MeV in the
centre-of-mass system, can be
described reasonably by the Powell
formula \cite{powell}. At higher energies $50<\omega<100$ MeV, the internal
structure of the system slowly becomes visible.
Historically this nucleon structure-dependent effect in {\it unpolarized}
Compton scattering was taken into account by introducing {\it two} free
parameters into the cross-section formula, commonly denoted the
{\it electric} $(\bar\alpha)$ and {\it magnetic} $(\bar\beta)$ polarizabilities of
the nucleon in analogy to the structure dependent response functions for
light-matter
interactions in classical electrodynamics. Over the past few decades several
experiments on low energy Compton scattering off the proton have taken place,
resulting
in several extractions of the electromagnetic polarizabilities of the proton.
At present,
the commonly accepted numbers are
$\bar\alpha^{(p)} = (12.1\pm 0.8\pm 0.5)\times 10^{-4}\,{\rm fm}^3$,
$\bar\beta^{(p)} = (2.1\mp 0.8\mp 0.5)\times 10^{-4}\,{\rm fm}^3$~\cite{1},
indicating that the proton compared to its volume of $\sim 1\,$fm$^3$ is 
a rather stiff object. In parallel to the ongoing experimental 
efforts theorists have tried
to understand the internal dynamics of the nucleon that would give rise to such
(small) structure effects. At present, several quite different theoretical
approaches
find qualitative and quantitative explanations for these 2 polarizabilities,
but it can be considered as one of the striking successes of chiral
perturbation theory\cite{ulfco} (for a general overview, see e.g. ref.\cite{CD}).

Quite recently, with the advent of polarized targets and new sources with a
high
flux of polarized photons, the case of {\it polarized} Compton scattering off
the
proton $\vec{\gamma}\,\vec{p}\rightarrow\gamma p$ has come close to
experimental
feasibility. On the theoretical side it has been shown \cite{ragusa} that one
can
define 4 spin-dependent electromagnetic response functions
$\gamma_i,\;i=1\dots 4$, which in analogy to $\bar{\alpha}\, , \bar{\beta}$
are
commonly called the ``spin-polarizabilites'' of the proton. First studies have
been published \cite{bab,gorchtein},
claiming that the such parameterized information on the low-energy spin
structure of the proton can really be extracted from the upcoming
double-polarization
Compton experiments. A success of this program would clearly shed new light on
our understanding of the internal dynamics of the proton and at the same time
serve
as a check on the theoretical explanations of the polarizabilities. The new
challenge to theorists will then be to explain all 6 of the leading
electromagnetic
response functions simultaneously. At present there only exists one
experimental
analysis that has shed some light on the magnitude of the (essentially) unknown
spin-polarizabilities $\gamma_i^{(p)}$ of the proton:
The LEGS group has reported \cite{legs} a result\footnote{Note that we have
subtracted
off the contribution of the pion-pole diagram in order to be consistent with
the
definition of the spin-polarizabilities given in \cite{hhkk}.} for a linear
combination involving three of the $\gamma_i$, namely
\begin{eqnarray}
\gamma_\pi^{(p)}|_{\rm exp.}
                   &=&\gamma_1^{(p)}+\gamma_2^{(p)}+2\gamma_4^{(p)}\nonumber
\\
                   &=&\left(17.3\pm3.4\right)\times10^{-4}\;{\rm fm}^4\;
                           \label{legsnumber}.
\end{eqnarray}
We note that this pioneering
result was obtained from an analysis of an {\em unpolarized} Compton
experiment in the backward direction, where the spin-polarizabilities come in
as
one contribution in a whole class of subleading order nucleon structure effects
in the
differential cross-section. Given these structure subtleties and the fact that
most
theoretical calculations \cite{bab,gorchtein,hhkk,Krein,nathan}
have predicted this particular linear combination of
spin-polarizabilities to be a factor of 2 smaller than the number given in
Eq.(\ref{legsnumber}), we can only reemphasize the need for the upcoming
polarized
Compton scattering experiments.

In this note we are taking up the challenge on the theory side
within the context of Heavy Baryon Chiral
Perturbation Theory (HBChPT), extending previous efforts
\cite{bkmrev,hhkk,Ji,Birse}
in a significant way. Previously an order ${\cal O}(p^3)$ SU(2) HBChPT
calculation
\cite{bkmrev} was performed, which showed that the leading (i.e. long-range)
structure effects in the spin-polarizabilities are given by 8 different $\pi N$
loop
diagrams giving rise to a $1/m_\pi^2$ behavior in the $\gamma_i$. Subsequently
it was
shown in an ${\cal O}(\epsilon^3)$ SU(2) ``small scale expansion''
(SSE) calculation \cite{hhkk}---which in
contrast
to HBChPT includes the first nucleon resonance $\Delta$(1232) as an explicit
degree
of freedom \cite{SSE}---that 2 ($\gamma_2,\;\gamma_4$) of the 4
spin-polarizabilities receive
large corrections due to $\Delta$(1232) related effects, resulting in a big
correction
to the leading $1/m_\pi^2$ behavior~\cite{saito}. 
Another important conclusion of
\cite{hhkk} was
that any HBChPT calculation that wants to calculate $\gamma_2,\;\gamma_4$ would
have
to be extended to ${\cal O}(p^5)$ before it can incorporate the large
$\Delta$(1232)
related corrections found in \cite{hhkk}. Recently, two ${\cal O}(p^4)$ SU(2)
HBChPT
calculations \cite{Ji,Birse} of polarized Compton scattering in the forward
direction
appeared, from which one can extract one particular
linear combination\footnote{$\gamma_0$ can also be calculated from the
absorption
cross sections of polarized photons on polarized nucleons via the GGT sum rule
\cite{GGT}, as pointed out in \cite{BKKM}. In the absence of such data several
groups
have tried
to extract the required cross sections via a partial wave analysis of
unpolarized
absorption cross sections. Recent results of these efforts are given in table
2.}
 of 3 of the 4 $\gamma_i$,
which is usually called $\gamma_0$:
\begin{eqnarray}
\gamma_0=\gamma_1-\left(\gamma_2+2\gamma_4\right)\cos\theta|_{\theta\rightarrow
0} \;.
\end{eqnarray}
The authors of \cite{Ji,Birse} claimed to have found a huge correction to
$\gamma_0$
at ${\cal O}(p^4)$ relative to the ${\cal O}(p^3)$ result already found in
\cite{BKKM},
casting doubt on the usefulness/convergence of HBChPT for
spin-polarizabilities. Given
that $\gamma_0$ involves the very 2 polarizabilities $\gamma_2,\gamma_4$, which
were
already shown in \cite{hhkk} to receive huge corrections even up to ${\cal
O}(p^5)$
when one tries to calculate them in an effective field theory without explicit
$\Delta$(1232) degrees of freedom, the (known) poor convergence for $\gamma_0$
found in \cite{Ji,Birse} should not have come as a surprise. We will come back
to this
point later.

In the following we report on the results of a ${\cal O}(p^4)$ calculation
of all 4 spin-polarizabilities $\gamma_i$, which allows to study the issue of
convergence
in chiral effective field theories for these important new spin-structure
parameters
of the nucleon. The pertinent results of our investigation can be summarized as
follows:\\
1) We first want to comment on the extraction of polarizabilities from nucleon
Compton
scattering amplitudes.
In previous analyses \cite{hhkk,bkmrev}
it has always been stated that in order to obtain the spin-polarizabilities
from the
calculated Compton amplitudes, one only has to subtract off
the nucleon tree-level graphs from the fully
calculated amplitudes. The remainder
in each (spin-amplitude) then started with a factor of $\omega^3$ and the
associated
Taylor-coefficient was related to the spin-polarizabilities. Due to the
(relatively)
simple structure of the spin-amplitudes at this order,
this prescription gives the correct result in the
${\cal O}(p^3)$ HBChPT \cite{bkmrev}
and the ${\cal O}(\epsilon^3)$ SSE \cite{hhkk} calculations. However, at
${\cal O}(p^4)$ (and also at ${\cal O}(\epsilon^4)$ \cite{ghm}) one has to
resort to a
definition of the (spin-) polarizabilities that is soundly based on
field theory,
in order to make sure that one
only picks up those contributions at $\omega^3$ that are really connected with
(spin-) polarizabilities. In fact, at ${\cal O}(p^4)$
(${\cal O}(\epsilon^4)$) the prescription given in \cite{hhkk,bkmrev} leads
to an
admixture of effects resulting from 2 successive, uncorrelated $\gamma NN$
interactions with a one nucleon intermediate state. In order to avoid these
problems we
advocate the following
definition for the {\em spin-dependent} polarizabilities in (chiral) effective
field theories:~{\em Given a complete set of spin-structure amplitudes
for Compton scattering to
a certain order in perturbation theory, one first removes all
one-particle (i.e. one-nucleon or one-pion) 
reducible (1PR) contributions from the full spin-structure amplitudes.}
Specifically, starting from the general form of the T-matrix for real Compton
scattering
assuming invariance under parity, charge conjugation and time reversal
symmetry, we utilize the following six structure amplitudes $A_i(\omega ,
\theta )$ \cite{hhkk,bkmrev} in the Coulomb gauge,
$\epsilon_0=\epsilon_0^\prime=0$,
\begin{eqnarray}
T &=& A_1(\omega,\theta)\vec{\epsilon}^{\, * \prime}\cdot\vec{\epsilon}
+A_2(\omega,\theta)\vec{\epsilon}^{\, * \prime}\cdot\hat{k} \; \vec{\epsilon}
\cdot\hat{k}^\prime \nonumber\\
&+&iA_3(\omega,\theta)\vec{\sigma}\cdot(\vec{\epsilon}^{\, * \prime}\times
\vec{\epsilon})
+iA_4(\omega,\theta)\vec{\sigma}\cdot(\hat{k}^\prime \times\hat{k})
\vec{\epsilon}^{\, * \prime} \cdot\vec{\epsilon} \nonumber\\
&+& iA_5(\omega,\theta)\vec{\sigma}\cdot[(\vec{\epsilon}^{\, * \prime} \times
\hat{k}) \vec{\epsilon}\cdot\hat{k}^\prime -(\vec{\epsilon}\times
\hat{k}^\prime ) \vec{\epsilon}^{\, * \prime} \cdot\hat{k}]\nonumber\\
&+& iA_6(\omega,\theta)\vec{\sigma}\cdot[(\vec{\epsilon}^{\, * \prime}\times
\hat{k}^\prime ) \hat{\epsilon}\cdot\hat{k}^\prime -(\vec{\epsilon}\times
\hat{k})\vec{\epsilon}^{\, * \prime} \cdot\hat{k}],
\end{eqnarray}
where $\theta$ corresponds to the c.m. scattering angle,
$\vec{\epsilon},\hat{k}\;
(\vec{\epsilon}^{\, \prime} ,\hat{k}^\prime )$ denote
the polarization vector, direction  of the incident (final) photon while
$\vec{\sigma}$ represents the (spin) polarization vector of the nucleon.
Each (spin-)structure amplitude is now separated into 1PR contributions and a
remainder,
that contains the response of the nucleon's excitation structure to two
photons:
\begin{eqnarray}
A_i(\omega,\theta)=A_i(\omega,\theta)^{\rm 1PR}+
A_i(\omega,\theta)^{\rm exc.} \, ,
i=3,\dots,6\, . \label{separation}
\end{eqnarray}
Taylor-expanding the spin-dependent
$A_i(\omega,\theta)^{\rm 1PR}$ for the case of a proton target in the c.m. frame
into
a power series in $\omega$, the leading terms are linear in $\omega$ and are
given by the venerable LETs of Low, Gell-Mann and Goldberger \cite{low}:
\begin{eqnarray}
A_3(\omega,\theta)^{\rm 1PR}&=&\frac{\left[1+2\kappa^{(p)}-(1+\kappa^{(p)})^2\cos
\theta\right]e^2}{2M_{N}^2}\,\omega+{\cal O}(\omega^2),\nonumber\\
A_4(\omega,\theta)^{\rm 1PR}&=&-{(1+\kappa^{(p)})^2e^2\over 2M_{N}^2}\,\omega
+{\cal O}(\omega^2),\nonumber\\
A_5(\omega,\theta)^{\rm 1PR}&=&{(1+\kappa^{(p)})^2e^2\over 2M_{N}^2}\,\omega
+{\cal O}(\omega^2),\nonumber\\
A_6(\omega,\theta)^{\rm 1PR}&=&-{(1+\kappa^{(p)})e^2\over 2M_{N}^2}\,\omega
+{\cal O}(\omega^2)~.
\end{eqnarray}
While it is not advisable to really perform this Taylor-expansion for the
spin-dependent $A_i(\omega,\theta)^{\rm 1PR}$
due to the complex pole structure, one can do so without problems
for the $A_i(\omega,\theta)^{\rm exc.}$ as long as $\omega \ll m_\pi$.
For the case of a proton one then finds
\begin{eqnarray}
A_3(\omega,\theta)^{\rm exc.}&=&4\pi \left[ \gamma_{1}^{(p)}-(\gamma_{2}^{(p)}+
2\gamma_{4}^{(p)}) \cos \theta \right]\omega^3+{\cal O}(\omega^4),\nonumber\\
A_4(\omega,\theta)^{\rm exc.}&=&4\pi\gamma_{2}^{(p)}\omega^3+{\cal
O}(\omega^4),\nonumber\\
A_5(\omega,\theta)^{\rm exc.}&=&4\pi\gamma_{4}^{(p)}\omega^3+{\cal
O}(\omega^4),\nonumber\\
A_6(\omega,\theta)^{\rm exc.}&=&4\pi\gamma_{3}^{(p)}\omega^3 +{\cal O}(\omega^4)
\label{xxx}\; .
\end{eqnarray}
We therefore take Eq.(\ref{xxx}) as starting point for the calculation of the
spin-polarizabilities, which are related to the $\omega^3$ Taylor-coefficients
of
$A_i(\omega,\theta)^{\rm exc.}$.
As noted above, both the ${\cal O}(p^3)$ HBChPT \cite{bkmrev}
and the ${\cal O}(\epsilon^3)$ SSE \cite{hhkk} results are consistent with this
definition. \\
2) Utilizing Eqs.(\ref{separation},\ref{xxx}) we
have calculated the first subleading correction, ${\cal O}(p^4)$, to
the 4 isoscalar spin-polarizabilities $\gamma_i^{(s)}$ already determined to
${\cal O}(p^3)$
in \cite{bkmrev} in SU(2) HBChPT.
We employ here the convention \cite{hhkk}
\begin{eqnarray}
\gamma_i^{(p)}=\gamma_i^{(s)}+\gamma_i^{(v)} \; ;\quad
\gamma_i^{(n)}=\gamma_i^{(s)}-\gamma_i^{(v)} \; .
\end{eqnarray}
Contrary to popular opinion we show, that even
at subleading order all 4  spin-polarizabilities can be given in closed form
expressions
which are free of any unknown chiral counterterms! The only parameters
appearing
in the results are the axial-vector nucleon coupling constant $g_A=1.26$, 
the pion decay
constant $F_\pi= 92.4\,$MeV, the pion mass $m_\pi= 138\,$MeV, 
the mass of the nucleon $M_N= 938\,$MeV as
well as
its isoscalar, $\kappa^{(s)} = -0.12$, and isovector, $\kappa^{(v)}=3.7$, 
anomalous magnetic moments.
All ${\cal O}(p^4)$ corrections arise from 25 one-loop $\pi N$ continuum
diagrams, with
the relevant vertices obtained from the well-known SU(2) HBChPT ${\cal O}(p)$
and
${\cal O}(p^2)$ Lagrangians given in detail in ref.\cite{bkmrev}.
To ${\cal O}(p^4)$ we find
\begin{eqnarray}
\gamma_1^{(s)}&=& + \; {e^2g_A^2\over 96\pi^3F_\pi^2m_\pi^2} 
                       \left[1-\mu \,\pi\right]~,\\
\gamma_2^{(s)}&=& + \; {e^2g_A^2\over 192\pi^3F_\pi^2m_\pi^2}  
                       \left[1+\mu \,\frac{(-6+\kappa^{(v)})\pi}{4}\right]~,\\
\gamma_3^{(s)}&=& + \; {e^2g_A^2\over 384\pi^3F_\pi^2m_\pi^2}  
                       \left[1-\mu \, \pi\right]~,\\
\gamma_4^{(s)}&=& - \; {e^2g_A^2\over 384\pi^3F_\pi^2m_\pi^2}  
                       \left[1-\mu \,\frac{11}{4}\pi\right]~, 
\end{eqnarray}
with $\mu = m_\pi / M_N \simeq 1/7$ and the
the numerical values given in table 1. The leading $1/m_\pi^2$ behavior
of the
isoscalar spin-polarizabilities is not touched by the ${\cal O}(p^4)$
correction,
as expected.
With the notable exception of $\gamma_4^{(s)}$,
which even changes its sign due to a large ${\cal O}(p^4)$ correction, we show
that
this first subleading order of $\gamma_1^{(s)},\gamma_2^{(s)},\gamma_3^{(s)}$
amounts
to a 25-45\% correction to the leading order result. This does not quite
correspond
to the expected $m_\pi/M_N$ correction of (naive) dimensional analysis, but can
be
considered acceptable. The physical origin of the large correction in
$\gamma_4^{(s)}$
is not yet understood, but we remind the reader again of our comments above,
that it
was shown in the SSE calculation of \cite{hhkk} that one should not expect a
good
convergence behavior for $\gamma_2^{(s)},\gamma_4^{(s)}$ in HBChPT at all.\\
3) We further report the first results for the 4 {\em isovector}
spin-polarizabilities
$\gamma_i^{(v)}$ obtained in the framework of chiral effective field theories.
Previous calculations at ${\cal O}(p^3)$ \cite{bkmrev} and
${\cal O}(\epsilon^3)$ \cite{hhkk} were only sensitive to the isoscalar
spin-polarizabilities $\gamma_i^{(s)}$, therefore this calculation gives the
first indication from a 
chiral effective field theory about the magnitude of the difference
in the low-energy spin structure between proton and neutron. As in the case of
the isoscalar spin-polarizabilities there are again no unknown counterterm
contributions
to this order in the $\gamma_i^{(v)}$. All ${\cal O}(p^4)$
contributions arise from 16 one-loop
$\pi N$ continuum diagrams with the relevant ${\cal O}(p),\;{\cal O}(p^2)$
vertices again obtained from the Lagrangians given in ref.\cite{bkmrev}.
To ${\cal O}(p^4)$ one finds
\begin{eqnarray}
\gamma_1^{(v)}&=&  {e^2g_A^2\over 96\pi^3F_\pi^2m_\pi^2} 
                       \left[0-\mu \, \frac{5\pi}{8}\right]~,\\
\gamma_2^{(v)}&=&  {e^2g_A^2\over 192\pi^3F_\pi^2m_\pi^2}  
                       \left[0-\mu \, \frac{(1+\kappa^{(s)})\pi}{4}\right]~,\\
\gamma_3^{(v)}&=&  {e^2g_A^2\over 384\pi^3F_\pi^2m_\pi^2}  
                       \left[0+\mu \, \frac{\pi}{4}\right]~,\\
\gamma_4^{(v)}&=&   0~, 
\end{eqnarray}
with the numerical values again given in table 1.
The result of our investigation is
that the size of the $\gamma_i^{(v)}$ really tends to be an order of
magnitude smaller than the one of the $\gamma_i^{(s)}$ (with the possible
exception
of $\gamma_1^{(v)}$), supporting the scaling expectation, $\gamma_i^{(v)}\sim
(m_\pi/M_N)
\gamma_i^{(s)}$ from (naive) dimensional analysis. This is reminiscent of the
situation
in the spin-independent electromagnetic polarizabilities
$\bar{\alpha}^{(v)},\bar{\beta}^{(v)}$ \cite{ulfco}, which are also
suppressed
by one chiral power relative to their isoscalar partners
$\bar{\alpha}^{(s)},\bar{\beta}^{(s)}$. \\
4) Finally, we want to comment on the comparison between our results and
existing
calculations using dispersion analyses. Given our comments on the convergence
of the
chiral expansion for the (isoscalar) spin-polarizabilities \cite{hhkk}, we
reiterate
that we do not believe our ${\cal O}(p^4)$ HBChPT result for
$\gamma_2^{(s)},\gamma_4^{(s)}$
to be meaningful. Their large inherent $\Delta$(1232) related contribution just
cannot be
included (via a counterterm) before ${\cal O}(p^5)$ in HBChPT that only deals
with pion
and nucleon degrees of freedom. In table 1 it is therefore
interesting to note that by adding (``by hand'') the delta-pole contribution of
$\sim-2.5\,10^{-4}$fm$^4$ found in \cite{hhkk} to $\gamma_2^{(s)}$ one could
get quite
close to the range for this spin-polarizability
as suggested by the dispersion analyses \cite{Krein,gorchtein,bab}. Similarly,
adding
$\sim+2.5\,10^{-4}$fm$^4$ to $\gamma_4^{(s)}$ as suggested by \cite{hhkk} also
leads
quite close to the range advocated by the dispersion results
\cite{Krein,gorchtein,bab}.
However, such a procedure is of course not legitimate in an effective field
theory,
but it raises the hope that an extension of the ${\cal O}(\epsilon^3)$ SSE
calculation of
\cite{hhkk} that includes explicit delta degrees of freedom could lead to a
much
better behaved perturbative expansion for the isoscalar spin-polarizabilities.
Whether
this expectation holds true will be known quite soon \cite{ghm}.
For the isovector spin-polarizabilities we have given the first predictions
available
from effective field theory. In general the agreement with the range advocated
by the
dispersion analyses is quite good. Judging from table 1 we note that the main
difference between the 2
analyses from Mainz \cite{Krein,gorchtein} seems to lie in the treatment of the
isovector structure, indicating that the isospin separation might pose some
difficulties in the dispersion approaches. 
In table 2 we give a comparison of our results for those linear combinations of
the
$\gamma_i$
that typically are the main focus of attention in the literature. However, we
re-emphasize
that we do not consider our ${\cal O}(p^4)$ HBChPT predictions for
$\gamma_0^{(s)}, \gamma_\pi^{(s)}$ to be meaningful, because they involve
$\gamma_2^{(s)},\gamma_4^{(s)}$. The corresponding isovector combinations,
however, again
seem to agree quite well with the dispersive results and so far we have no
reason
to suspect that they might be affected by the poor convergence behavior of
some of their
isoscalar counterparts. We further note that our ${\cal O}(p^4)$ HBChPT
predictions for $\gamma_0^{(s,v)}$ differ from the ones given in 2 recent
calculations
\cite{Ji,Birse}. As noted above this difference solely arises from a different
definition
of nucleon spin-polarizabilities. If we (``by hand'')
Taylor-expand our $\gamma NN$ vertex
functions in powers of $\omega$ and include the resulting terms into the the
$\gamma_0$
structure, we obtain the ${\cal O}(p^4)$ corrections $\gamma_0^{(s)}=-6.9,\;
\gamma_0^{(v)}=-1.6$ in units of $10^{-4}$fm$^4$, in numerical (and analytical)
agreement
with \cite{Ji,Birse}. This brings us to an important point: Once the first
polarized
Compton asymmetries have been measured, it has to be checked very carefully
whether the
same input data fitted to the terms we define as 1PR plus the additional free
$\gamma_i$
parameters leads to the same numerical fit-results for the
spin-polarizabilities as
in the dispersion theoretical codes usually employed to extract polarizabilities
from
Compton data. Small
differences for example in the treatment of the pion/nucleon pole could lead to
quite
large systematic errors in the determination of the $\gamma_i$. Such studies
are under
way \cite{ghm}.

\acknowledgements

G.C.G. would like to acknowledge financial support from the TMR network HaPHEEP
under contract FMRX-CT96-0008.

\begin{table}
\begin{tabular}{c||cc|c||cccc}
$\gamma_{i}^{(N)}$ & ${\cal O}(p^3)$ & ${\cal O}(p^4)$ & Sum  & Mainz1 & Mainz2
&
BGLMN & SSE1  \\
\hline
$\gamma_{1}^{(s)}$ & $+ 4.6$ & $-2.1$ & $+2.5$ & $+5.6$ &$+5.7$ &  $+4.7$
&$+4.4$\\
$\gamma_{2}^{(s)}$ & $+ 2.3$ & $-0.6$ & $+1.7$ & $-1.0$ &$-0.7$ &  $-0.9$
&$-0.4$\\
$\gamma_{3}^{(s)}$ & $+ 1.1$ & $-0.5$ & $+0.6$ & $-0.6$ &$-0.5$ &  $-0.2$
&$+1.0$\\
$\gamma_{4}^{(s)}$ & $- 1.1$ & $+1.5$ & $+0.4$ & $+3.4$ &$+3.4$ &  $+3.3$
&$+1.4$\\
\hline
$\gamma_{1}^{(v)}$ & - & $-1.3$ & $-1.3$ & $-0.5$ & $-1.3$ & $-1.6$ & - \\
$\gamma_{2}^{(v)}$ & - & $-0.2$ & $-0.2$ & $-0.2$ & $+0.0$ & $+0.1$ & - \\
$\gamma_{3}^{(v)}$ & - & $+0.1$ & $+0.1$ & $-0.0$ & $+0.5$ & $+0.5$ & - \\
$\gamma_{4}^{(v)}$ & - & $+0.0$ & $+0.0$ & $+0.0$ & $-0.5$ & $-0.6$ & - \\
\hline
\end{tabular}
\bigskip
\caption{\label{t1}
Predictions for the spin-polarizabilities in HBChPT in comparison with the
dispersion
analyses of refs.\protect \cite{Krein,gorchtein,bab} (Mainz1,Mainz2,BGLMN) and
the
${\cal O}(\epsilon^3)$ results of the
small scale expansion \protect \cite{hhkk} (SSE1).
All results are given in the units of $10^{-4}\;{\rm fm}^4$.}
\end{table}

\begin{table}
\begin{tabular}{c||cc|c||cccc}
$\gamma_{i}^{(N)}$ & ${\cal O}(p^3)$ & ${\cal O}(p^4)$ & Sum  & Mainz1 & Mainz2
&
BGLMN & SSE1  \\
\hline
$\gamma_{0}^{(s)}$   & $+ 4.6$ & $-4.5$ & $+0.1$ & $-0.2$  &$-0.4$  & $-1.0$&
+2.0 \\
$\gamma_{0}^{(v)}$   & -       & $-1.1$ & $-1.1$ & $-0.3$  &$-0.4$  & $-0.5$& -
\\
$\gamma_{\pi}^{(s)}$ & $+ 4.6$ & $+0.3$ & $+4.9$ & $+11.4$ &$+11.8$ & $+10.4$&
+6.8\\
$\gamma_{\pi}^{(v)}$ & -       & $-1.5$ & $-1.5$ & $-0.7$  &$-2.4$  &  $-2.7$&
- \\
\end{tabular}
\bigskip
\caption{\label{t2}
Predictions for the so-called forward (backward) spin-polarizabilities
$\gamma_0$
($\gamma_\pi$). For a definition of units and references see table 1.}
\end{table}

\end{document}